\documentclass[twocolumn,superscriptaddress,aps,prb]{revtex4-2}
\usepackage{CJKutf8}
\usepackage{bm}
\usepackage{appendix}
\usepackage{graphicx}
\usepackage{amsmath}
\usepackage{amssymb}%
\usepackage{color}
\usepackage{lipsum}
\usepackage{multirow}
\usepackage[colorlinks=true,citecolor=blue,urlcolor=blue]{hyperref}
\usepackage{mathrsfs}
\usepackage{braket}
\usepackage{hhline}
\usepackage{ulem}
\usepackage{hyperref}

\begin{document}
\begin{CJK*}{UTF8}{gbsn}
		
   \title{Liouvillian skin effect in a one-dimensional open many-body quantum system with generalized boundary conditions}
	
	\author{Liang Mao}
	\affiliation{Institute for Advanced Study, Tsinghua University, Beijing, 100084, China}
		\author{Xuanpu Yang}
	\affiliation{School of Physics, Nankai University, Tianjin 300071, China}
	\author{Ming-Jie Tao}
	\email{MJT@iphy.ac.cn}
	\affiliation{College of Mathematics and Physics, Chengdu University of Technology, Chengdu 610059, China}
		\author{Haiping Hu}
		\email{hhu@iphy.ac.cn}
	\affiliation{Beijing National Laboratory for Condensed Matter Physics, Institute of Physics, Chinese Academy of Sciences, Beijing 100190, China}
		\affiliation{School of Physical Sciences, University of Chinese Academy of Sciences, Beijing 100049, China}
	\author{Lei Pan}
	\email{panlei@nankai.edu.cn}
	\affiliation{School of Physics, Nankai University, Tianjin 300071, China}

\begin{abstract}
Non-Hermitian skin effect (NHSE), namely that eigenstates of non-Hermitian Hamiltonains are localized at one boundary in the open boundary condition, attracts great interest recently.
In this paper, we investigate the skin effect in one-dimensional dissipative quantum many-body systems, which we call the Liouvillian skin effect (LSE). We rigorously identify the existence of LSE for generalized boundary conditions by solving the Liouvillian superoperator of an exactly solvable model with the advantage of  Bethe ansatz. The LSE is sensitive to boundary conditions where the signature is reflected in eigenfunctions of the system. We confirm that the LSE is fragile to a tiny co-flow boundary hopping with non-Hermitian current but can survive for a counter-flow boundary hopping in the thermodynamic limit.  
Our work provides a prototypical example of exactly solvable dissipative quantum many-body lattice systems exhibiting LSE for generalized boundary conditions. It can be further extended to other integrable open quantum many-body models. 
	\end{abstract}
	
	\maketitle
\end{CJK*}	
	
	
\section{Introduction}
Open systems are ubiquitous, covering the atomic realm to the observable universe. 
In recent years, open quantum many-body physics has become an important area \cite{NH_Review1,NH_Review2,NH_Review3} principally due to experimental advances in accurately controlling dissipations and interactions between particles \cite{Exp1,Exp2,Exp3,Exp4,Exp5,Exp6,Exp7,Exp8,Exp9,Exp10,Exp11,Exp12} and has received considerable theoretical attention \cite{NHMB1,NHMB2,NHMB3,NHMB4,NHMB405,NHMB5,NHMB6,NHMB7,NHMB8,NHMB801,NHMB9,NHMB11,NHMB12,NHMB125,NHMB126,NHMB13,NHMB14,NHMB15,NHMB16,NHMB17,NHMB18,NHMB19,NHMB20,NHMB21,NHMB22,NHMB23,NHMB24,NHMB25,NHMB26,NHMB27,NHMB28,NHMB29,NHMB30,NHMB301,NHMB302,NHMB303,NHMB304,NHMB31,NHMB32,NHMB33,NHMB34,NHMB35,NHMB355,NHMB356,NHMB357,NHMB358,NHMB36,NHMB37,NHMB38,NHMB39,NHMB40,NHMB41,NHMB42,NHMB43,NHMB44,NHMB45,NHMB46,NHMB47,NHMB48,NHMB49,NHMB50,NHMB51,NHMB52,NHMB53}. Furthermore, recent theoretical progress has revealed that many rich phenomena can emerge from the interplay between dissipations and interparticle interactions, including but not limited to parity-time symmetric quantum criticality \cite{NHMB6,NHMB7}, anomalous non-exponential dynamics \cite{NHMB8}  and non-Hermitian many-body localization \cite{NHMB35,NHMB355,NHMB356,NHMB357,NHMB358}, negative central charge at an exceptional point \cite{NHMB49,NHMB50} et cetera.

A quantum system coupled to environment degrees of freedom constitutes an open quantum system whose time-evolution is described by a Lindblad master equation under the Markovian approximation. The dynamics of an open quantum system are dominated by the Liouvillian superoperator and could be described by the effective non-Hermitian Hamiltonian under some circumstances. 
It has been shown that the non-Hermitian Hamiltonian exhibits unique features, among which the non-Hermitian skin effect (NHSE) \cite{Skin_Theory1,Skin_Theory2,Skin_Theory201,Skin_Theory3,Skin_Theory4} attracts a growing attention both  theoretically \cite{Skin_Theory5,Skin_Theory6,Skin_Theory7,Skin_Theory8,Skin_Theory9,Skin_Theory10,Skin_Theory11,Skin_Theory12,Skin_Theory125,Skin_Theory13,Skin_Theory131,Skin_Theory14,Skin_Theory15,Skin_Theory16,Skin_Theory17,Skin_Theory18,Skin_Theory185,Skin_Theory19,Skin_Theory20,Skin_Theory21,Skin_Theory22,Skin_Theory23,Skin_Theory24,Skin_Theory25} and experimentally \cite{Skin_Exp1,Skin_Exp2,Skin_Exp3,Skin_Exp4,Skin_Exp5,Skin_Exp6,Skin_Exp7,Skin_Exp8,Skin_Exp9}. 
The NHSE states that the eigenstates of non-Hermitian Hamiltonian can be localized at the boundaries in the open boundary conditions, albeit remain extended in the periodic boundary condition.
This effect originates from the sensitivity of related non-Hermitian terms in the Hamiltonian against boundary perturbation, and manifest itself in not only eigenstates properties, but also in distinct features of eigenvalues under different boundary conditions.

Previous investigations primarily dealt with the NHSE at the level of single-particle systems and revealed that the NHSE has a dramatic influence on the topology and dynamics.  Recently, a number of studies have begun to examine the NHSE in the framework of many-body systems \cite{Many_Skin_Theory1,Many_Skin_Theory2,Many_Skin_Theory3,Many_Skin_Theory4,Many_Skin_Theory5,Many_Skin_Theory6,Many_Skin_Theory7,Many_Skin_Theory8,Many_Skin_Theory9,Many_Skin_Theory10,Many_Skin_Theory11,Many_Skin_Theory12,Many_Skin_Theory13,Many_Skin_Theory14,Many_Skin_Theory14,Many_Skin_Theory15,Many_Skin_Theory16,Many_Skin_Theory17} and found that the interplay between the NHSE and interaction can give rise to impact on the properties such as slowdown of relaxation dynamics \cite{Many_Skin_Theory11}, many-body localization \cite{Many_Skin_Theory11,Many_Skin_Theory12} and entanglement transition \cite{Many_Skin_Theory13,Many_Skin_Theory14,Many_Skin_Theory15}.
More recently, the skin effect has been identified to exist in the Lioivillian superoperator due to its intrinsic non-Hermiticity, which was named the Liouvillian skin effect (LSE) \cite{LSE1,LSE2,LSE3,LSE4}. Meanwhile, the research on exactly solvable models of open quantum many-body systems gradually increased where the structure of Liouvillian in the framework of Yang-Baxter integrability is a new emerging field of mathematical physics \cite{Prosen2016,Prosen2019,Essler2020,EsslerSci,Leeuw2021,Claeys2022,Martin2022,Popkov2021,Prosen2022,Prosen2022_2,WangzhongScipost,NH_BHM}. 

In the literature, Guo \textit{et al}. \cite{Skin_Theory6} demonstrated the existence condition of the NHSE for more general boundary conditions at the single-particle level. Moreover, recent study \cite{NH_BHM} reveals that boundary conditions have a significant influence on the integrability of non-Hermitian systems. It is natural to begin investigating the existence and reveal the connection between the LSE and boundary conditions at the level of many-body physics. 
In this paper, we explore the relationship between LSE and different boundary conditions by means of techniques to solve the Liouvillian superoperators that were developed recently. By constructing an exactly solvable Liouvillian superoperator,  we find the exact solutions for more general boundary conditions, allowing us to unearth the evidence for existence of the LSE. 
We find that the LSE will be destroyed by a certain type of boundary perturbation but immune to another type of perturbation. Our work provides a firm ground for the existence of LSE and its sensitivity to boundary effects from the exactly solvable many-body systems perspective.

The rest of this paper is organized as follows. In Sec. \ref{Sec. Model}, we introduce an exactly solvable Liouvillian superoperator that can be mapped to a non-Hermitian XXZ chain in the subspace.
Sec. \ref{Sec. PBC & OBC} solves the exact solution in the PBC and OBC. Then, we discuss the general boundary condition and identify the existence condition of the LSE in Sec. \ref{Sec. GBC}. Finally, we summarize in Sec. \ref{Summary}. All the calculation details are provided in the appendices.   

	
\begin{figure}[!t]
	\includegraphics[width=8.5cm]{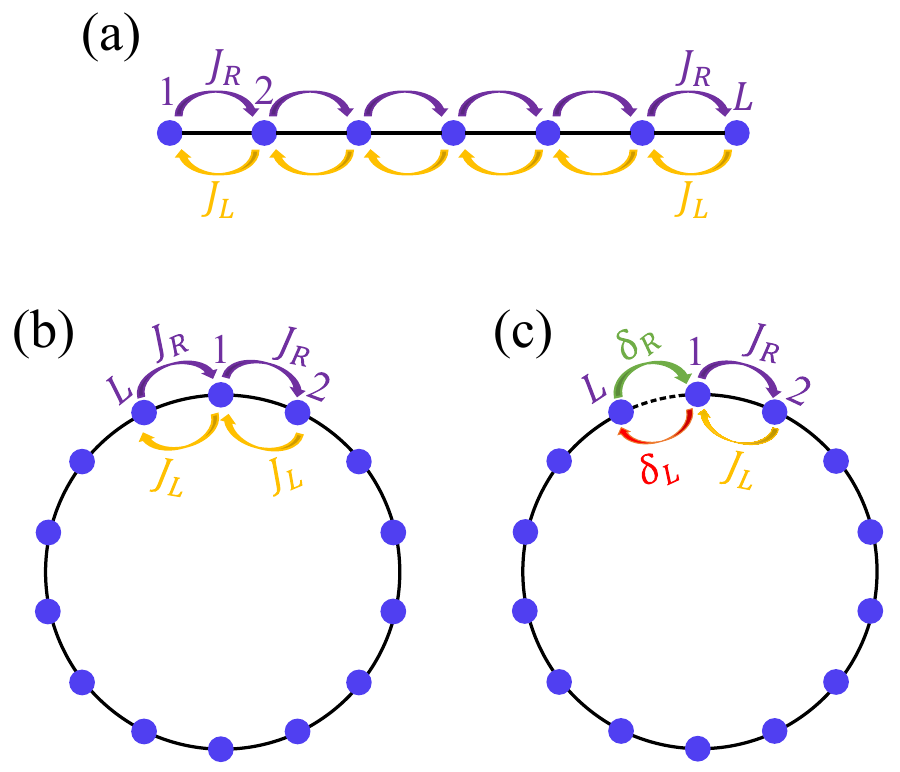}
	\caption{Illustration of three boundary conditions. (a) The Open boundary condition (OBC) where no hopping between two edges. (b) The periodic boundary condition (PBC). (c) The generalized boundary condition (GBC) where the hopping amplitudes are unequal to ones in the bulk ($\delta_L\neq J_L$, $\delta_R\neq J_R$.).}
	\label{Fig1}
\end{figure}
\section{The Model}
\label{Sec. Model}
Time evolution of the density matrix $\rho(t)$ for a Markovian open quantum system is generally governed by the Lindblad master equation \cite{Lindblad1,Lindblad2} ($\hbar=1$), 
\begin{align}
	\frac{d \rho(t)}{d t}  \equiv \hat{\mathscr{L}} \rho(t) =-i\left[\hat{H}_{S}, \rho(t)\right]+\mathcal{D}[\rho(t)],\label{LME} 
\end{align}
where $\hat{\mathscr{L}}$ is named Liouvillian superoperator. The first term corresponds to the coherent evolution and the second term  $\mathcal{D}[\rho(t)]=\sum_k\Big(2 \hat{L}_k \rho(t) \hat{L}^{\dagger}_{k}-\left\{\hat{\rho}(t), \hat{L}^{\dagger}_{k} \hat{L}_k\right\}\Big)$ gives the dissipation process.  Here $[\cdot,\cdot]$ and $\{\cdot,\cdot\}$ denote commutator and anticommutator, respectively.
According to the Choi-Jamiołkowski isomorphism \cite{CJ1,CJ2}, the Lindblad equation can be expressed as equivalent equation $\frac{d}{dt}|\rho\rangle=\hat{\mathscr{L}}|\rho\rangle$ with vectorized density matrix $|\rho\rangle=\sum_{i, j} \rho_{i, j}|i\rangle \otimes|j\rangle$ in double space $\mathcal{H}_R\otimes\mathcal{H}_L$, and the Liouvillian superoperator is written as
\begin{align}
	\hat{\mathscr{L}}= & -i\left(\hat{H}_{S} \otimes \hat{I}-\hat{I} \otimes \hat{H}_{S}^{\mathrm{T}}\right) \nonumber \\
	& +\sum_k\left[2 \hat{L}_k \otimes \hat{L}^{*}_{k}-\hat{L}^{\dagger}_{k} \hat{L}_{k} \otimes \hat{I}-\hat{I} \otimes\left(\hat{L}^{\dagger}_{k} \hat{L}_{k}\right)^{\mathrm{T}}\right].
\end{align}

The Lindblad equation is determined by the spectrum of $\hat{\mathscr{L}}$. The system reaches a steady state which is the zero-energy eigenstate of $\hat{\mathscr{L}}$.
The solvability of $\hat{\mathscr{L}}$ is clearly involved the choice of operators $\hat{L}_{k}$. 
Typically, the Liouvillian superoperator will almost never be integrable unless special circumstances; for example $\hat{L}_k$ is chosen as
the single creation and annihilation operator and meanwhile the Hamiltonian $\hat{H}_{S}$ is simply integrable model. 
Ref. \cite{Essler2020} found an alternative mechanism connecting the Lindblad superoperator to known integrable models. Suppose there exists local projectors $\mathcal{\hat{P}}_{j}^{m}$ acting on the double space which satisfy $\mathcal{\hat{P}}_{j}^{m}\mathcal{\hat{P}}_{j}^{n}=\delta_{m,n}\mathcal{\hat{P}}_{j}^{m}$ and  $\mathcal{\hat{P}}_{j}^{0}=1-\sum_m\mathcal{\hat{P}}_{j}^{m}$. If all projectors commute with the $\hat{\mathscr{L}}$, i.e.,   $\Big[\hat{\mathcal{L}},\mathcal{\hat{P}}_{j}^{m}\Big]=0$, then the double space can be
decomposed by a series of subspaces which are eigenspace of projectors set $\{\hat{P}_{j}^{m}\}$.  These subspaces are invariant spaces of the Liouvillian superoperator. If we choose the dissipation operator $\hat{L}_{k}$ and Hamiltonian $\hat{H}_{S}$ such that Liouvillian superoperator in projected subspaces matches integrable models, the full spectrum of $\hat{\mathscr{L}}$ can be obtained analytically by solving each subspace individually.

Based on the above consideration, we consider a translational invariant one-dimensional (1D) spin chain with length $L$ and the dissipation acts on the system locally. then the Liouvillian superoperator is rewritten as
\begin{align}
	\hat{\mathscr{L}}=& -i\left(\hat{H}_{S} \otimes \hat{I}-\hat{I} \otimes \hat{H}_{S}^{\mathrm{T}}\right) +\sum_{j=1}^{L}\sum_{n=1}^{M}\Bigg[2 \hat{L}_{j}^{(n)} \otimes \hat{L}^{*(n)}_{j}\nonumber \\
	& -\hat{L}^{\dagger(n)}_{j} \hat{L}_{j}^{(n)} \otimes \hat{I}-\hat{I} \otimes\left(\hat{L}^{\dagger(n)}_{j} \hat{L}_{j}^{(n)}\right)^{\mathrm{T}}\Bigg]. \label{L_M}
\end{align}
where $j$ is the site index, and $M$ denotes the number of dissipation channels on each site. 
In order to derive an exactly solvable Liouvillian superoperator, the choice of operators $\hat{L}_{j}^{(n)}$ and Hamiltonian $\hat{H}_{S}$ should guarantee the integrability of $\hat{\mathscr{L}}$. We first focus on purely dissipative case, i.e., $\hat{H}_S=0$.
 We choose the local dissipation channels as  $\hat{L}_{j}^{(1)}=\sqrt{J_L}\hat{S}_{j}^+\hat{S}^-_{j+1}$ and $\hat{L}_{j}^{(2)}=\sqrt{J_R}\hat{S}_{j+1}^+\hat{S}_{j}^-$, where $S_j^{\pm}$ is the local spin operator.  Defining local projectors $\mathcal{\hat{P}}_{j}^{1}=\ket{1}^R_j\ket{0}^L_j\bra{0}^L_j\bra{1}^R_j,\mathcal{\hat{P}}_{j}^{2}=\ket{0}^R_j\ket{1}^L_j\bra{1}^L_j\bra{0}^R_j$ and  $\mathcal{\hat{P}}_{j}^{0}=1-\mathcal{\hat{P}}_{j}^{1}-\mathcal{\hat{P}}_{j}^{2}$, where $\ket{1}$ and $\ket{0}$ denotes spin up and down, respectively.$R$ and $L$ operators acting on the right and left Hilbert spaces  One can find easily
$\mathcal{\hat{P}}_{j}^{k}\mathcal{\hat{P}}_{j}^{l}=\delta_{k,l}\mathcal{\hat{P}}_{j}^{k}$ and the Liouvillian superoperator commutes with these projectors. 

Next we consider the effective Liouvillian superoperator in the projected subspaces. 
The most representative subspace is the diagonal subspace 
$\frac{d}{dt}\mathcal{\hat{P}}^{0}|\rho\rangle=\mathcal{\hat{P}}^{0}\hat{\mathscr{L}}|\rho\rangle=\hat{\mathscr{L}}_{\rm eff} \mathcal{\hat{P}}^0|\rho\rangle$ where $\mathcal{\hat{P}}^{0}=\prod_j\mathcal{\hat{P}}^{0}_{j}$ and $\mathscr{\hat{L}}_{\rm eff}=\mathcal{\hat{P}}^0\mathscr{\hat{L}}\mathcal{\hat{P}}^0$. 
We now derive the effective Liouvillian $\mathscr{\hat{L}}_{\rm eff}$. First we derive the effective action of individual $\hat{\mathscr{L}}$ terms
\begin{align}\label{Leff}
	\hat{\mathscr{L}}^{(1,2)}_{j, \rm eff}&=
	\mathcal{\hat{P}}^0\bigg[
	2\hat{L}_j^{(1,2)}\otimes \hat{L}^{*(1,2)}_j
	-\hat{L}^{\dagger (1,2)}_j\hat{L}^{(1,2)}_j\otimes\hat{I}\notag\\
	&\quad\quad\quad-\hat{I}\otimes\Big(
	\hat{L}^{\dagger (1,2)}_j\hat{L}^{(1,2)}_j
	\Big)^T
	\bigg]\mathcal{\hat{P}}^0\notag\\
	&=J_{L,R}\bigg[\hat{S}_j^+\hat{S}_{j+1}^-
	+\Big(
	\hat{S}_j^z\hat{S}_{j+1}^z-\frac{1}{4}
	\Big)\notag\\
	&\quad\quad\quad+\frac{1}{2}\Big(\mp\hat{S}_j^z\pm\hat{S}_{j+1}^z\Big).
	\bigg] 
\end{align}
From the Liouvillian  \eqref{Leff}, one can see clearly that non-reciprocal 
hopping derives from unequal dissipation strengths of $\hat{L}_{j}^{(1)}$ and $\hat{L}_{j}^{(2)}$. When adding $\hat{\mathscr{L}}^{(1)}_{j,\rm eff}$ or $\hat{L}^{(2)}_{j,\rm eff}$ terms together, PBC ensures that the contribution of the last term vanishes. Further taking into account both $\hat{\mathscr{L}}^{(1,2)}_{j,\rm eff}$ terms, we arrive at the effective Liouvillian
\begin{align} 
	\hat{\mathscr{L}}_{\rm eff}=2 J \sum_{j=1}^{L}& \Big[\frac{e^{-\phi}}{2} \hat{S}_j^{+} \hat{S}_{j+1}^{-}+\frac{e^\phi}{2} \hat{S}_{j+1}^{+} \hat{S}_j^{-}\nonumber \\
	&+\cosh \phi\left(\hat{S}_j^z \hat{S}_{j+1}^z-\frac{1}{4}\right)\Big].  \label{xxz}
\end{align}
Here we reset parameters $J_L,J_R$ via $J,\phi$ ($J_L=Je^{-\phi}$, $J_R=Je^{\phi}$). 
The original parameters $J_R$ ($J_L$) denotes the strength of up-spin hops to right (left) and the PBC is adopted as schematically displayed in Fig. \ref{Fig1}. We note that this model can be mapped to the Hatano-Nelson model \cite{HN_Model} with nearest-neighbor interaction by a Jordan-Wigner transformation.
 Without loss of generality,  we assume $\phi>0$ which means the right hopping is greater than the left. The solution of $\phi<0$ is connected to the $\phi>0$ case by inversion transformation $(\hat{S}_j^{+},\hat{S}_j^{-},\hat{S}_j^{z})\rightarrow(\hat{S}_{L+1-j}^{+},\hat{S}_{L+1-j}^{-},\hat{S}_{L+1-j}^{z})$. This model will be our main focus in the following discussion. A similar asymmetric non-Hermitian XXZ model is found exactly solvable in the PBC \cite{NH_XXZ,NH_Hubbard}.

Moving on now to consider other boundary conditions. For the OBC as shown in Fig. \ref{Fig1}(a), we can derive the following with same procedure as the PBC case. Following Eq. \eqref{Leff}, we now have single $\hat{S}^z$ terms dangling at each end of the boundary:
\begin{align}
	\hat{\mathscr{L}}_{\rm eff}^{\text {OBC}}&=J \sum_{j=1}^{L-1}\Bigg[e^{\phi} \hat{S}_{j+1}^{+} \hat{S}_{j}^{-}+e^{-\phi} \hat{S}_{j}^{+} \hat{S}_{j+1}^{-}\nonumber \\
	&+2 \cosh \phi\left(\hat{S}_{j}^{z} \hat{S}_{j+1}^{z}-\frac{1}{4}\right)\Bigg]+J\sinh \phi\Big(\hat{S}_{L}^{z}- \hat{S}_{1}^{z}\Big).\label{L_OBC}
\end{align}
The effective model \eqref{L_OBC} is also exactly solvable since boundary terms preserve the $U(1)$ symmetry. Moreover, we further consider generalized boundary conditions (GBCs). The effective Liouvillian is derived by adding extra boundary terms from Eq. \eqref{Leff} to the OBC effective Liouvillian Eq. \eqref{L_OBC}

\begin{align}
	\hat{\mathscr{L}}_{\rm eff}^{\text {GBC}}&=J \sum_{j=1}^{L-1}\Bigg[e^{\phi} \hat{S}_{j+1}^{+} \hat{S}_{j}^{-}+e^{-\phi} \hat{S}_{j}^{+} \hat{S}_{j+1}^{-}\nonumber \\
	&+2 \cosh \phi\left(\hat{S}_{j}^{z} \hat{S}_{j+1}^{z}-\frac{1}{4}\right)\Bigg]+J\sinh \phi\Big(\hat{S}_{L}^{z}- \hat{S}_{1}^{z}\Big)\nonumber\\
	&+\delta_L\left[\hat{S}_{L}^{+} \hat{S}_{1}^{-}
	+\left(\hat{S}_{L}^{z}-\frac{1}{2}\right) \left(\hat{S}_{1}^{z}+\frac{1}{2}\right)\right]\nonumber\\
	&+\delta_R\left[\hat{S}_{1}^{+} \hat{S}_{L}^{-}
	+\left(\hat{S}_{1}^{z}-\frac{1}{2}\right) \left(\hat{S}_{L}^{z}+\frac{1}{2}\right)\right],
	\label{L_GBZ1}
\end{align}
with boundary couplings $\delta_L, \delta_R$. The Liouvillian superoperator \eqref{L_GBZ1} is negative semidefinite whose zero-energy eigenstate corresponds to the steady state. The OBC case is $\delta_L=\delta_R=0$, and when $\delta_L=Je^{-\phi}, \delta_R=Je^{\phi}$ the GBC goes back to the PBC case. Exploring the physical properties of the system in the GBC takes on special significance. 

Before proceeding, we now discuss the effect of system Hamiltonian part and other projected subspaces. The system Hamiltonian $\hat{H}_S$, as demonstrate above, should be choose to ensure the integrability in projected subspaces. In the guidance of this principle, the integrability will be protected as long as the added Hamiltonian preserves the projected subspaces invariant and meanwhile does not destroy the dissipation part. It is straight to write down a generic Hamiltonian for nearest-neighbor form $\hat{H}_S=\sum_{j=1}^{L} J^{\prime} \hat{S}_{j}^{z} \hat{S}_{j+1}^{z}+h \hat{S}_{j}^{z}$. The coherent term $\hat{H}_{S} \otimes \hat{I}-\hat{I} \otimes \hat{H}_{S}^{\mathrm{T}}$ is diagonal in different invariant subspaces since $\hat{S}_z$ dose not flip spin on each site, the whole Liouvillian superoperator maintains its integrability.
Under such circumstance, we then ignore coherent term because it only add a  diagonal constant and  has no influence on the subsequent discussion about LSE.  
For other projected subspaces ($\mathcal{\hat{P}}\neq\mathcal{\hat{P}}^{0}$), the effective Liouvillians superoperators are connected non-Hermitian XXZ chains with different parameters \cite{Essler2020}. So in the rest of papers, we will only study the non-Hermitian XXZ model defined in Eq.\ref{xxz} through Eq.\ref{L_GBZ1}.

\section{Solution in the periodic and open boundary conditions}
\label{Sec. PBC & OBC}
This section will examine exact solutions of the Liouvillian superoperator in the PBC and OBC. 
Detailed analytical derivations are shown in the appendix \ref{App1} and \ref{App2}. In the following we outline the main results.

We first deal with the PBC case. Since the number of up-spin particle $N_\uparrow$ or down-spin particle $N_\downarrow$ is conserved quantity, we can solve the non-Hermitian XXZ model individually in different subspace with fixed $(N_\uparrow,N_\downarrow)$
\begin{align} 
	|\psi\rangle=\sum_{j=1}^M \sum_{x_j=1}^L \varphi\left(x_1, x_2, \cdots, x_M\right) S_{x_1}^{+} S_{x_2}^{+} \cdots S_{x_M}^{+}|\mathrm{vac}\rangle, \label{Wavefun_PBC1}
\end{align} 
where $|\operatorname{vac}\rangle=|\downarrow \downarrow \cdots \cdot \downarrow\rangle$ denotes the vacuum (no boson excited) state and $M=N_\uparrow$. The wavefunction $\varphi\left(x_1, x_2, \cdots, x_M\right)$ is expressed by the Bethe ansatz form
\begin{align} 
	\varphi\left(x_1, x_2, \cdots, x_M\right)=\sum_{\bf P} A_{\bf P} \exp \left(i \sum_{j=1}^M k_{p_j} x_j\right),
\end{align}
where $\{k_j\}$ is quasimomenta and the eigen-spectrum (dispersion relation) is determined with respect to the quasimomentum

\begin{align} 
	E=E_0+2J \sum_{j=1}^M\left[\cos \left(k_j+i\phi\right)-\cosh \phi\right],
\end{align} 
where the quasimomenta are determined by the Bethe ansatz equations (BAEs) \eqref{App_BAEs1}. By solving the BAEs, we get gapless excitation expectrum, resembling the behavior of Hermitian XXZ model in the gapless phase.

Actually, it was pointed out in  Ref \cite{NH_XXZ} that if one considers an artificial non-Hermitian XXZ model with tunable $S^z$ coupling, the model undergoes a phase transition from gapless to gapped phases by increasing $\phi$ while maintaining the $S^z$ coupling. However, the physical effective Liouvillian considered in our paper has the property that magnon hopping strength is related to $S^z$ coupling, so the system remains in the gapless phase for any non-zero $\phi$. This means that the non-Hermitian XXZ model considered here is in the same phase with the Hermitian gapless phase. Their ground states thus have no qualitative difference. The difference between Hermitian and non-Hermitian models is only that the latter one has complex spectrum.

\begin{figure}[!t]
	\includegraphics[width=8.5cm]{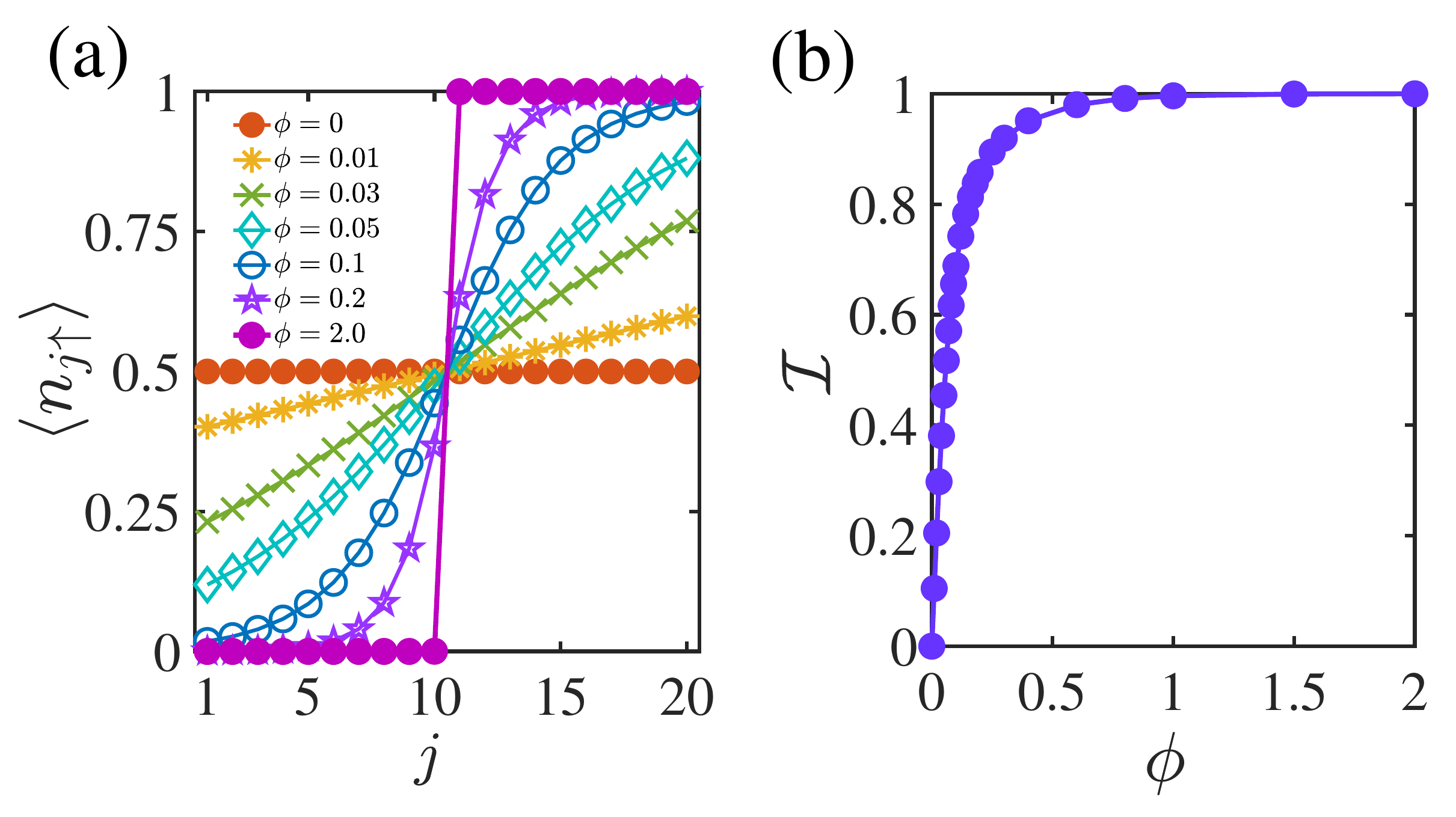}
	\caption{LSE in the OBC in the half filling. (a) The occupation number of the spin-up at each site for different $\phi$. (b)  Spin imbalance $\mathcal{I}$ as function of $\phi$.  Here we choose the system length $L=20$, and  the number of up-spins $M=10$.}
	\label{Fig2}
\end{figure}

For the OBC case, the non-Hermitian XXZ model \eqref{L_OBC} can also be  exactly solved as long as a matched ansatz is chosen. Starting from the ansatz
\begin{align}
 \varphi\left(x_1, x_2, \cdots, x_M\right)&=\sum_{\bf P, r_1,\cdots,r_M} A_{\bf P}(r_1,r_2,\cdots,r_M)\notag\\
 &\times \exp \left[\sum_{j=1}^M\left(i r_ j k_{p_j} x_j+\phi x_j\right)\right],
 \end{align}
 where $r_j=\pm 1$ denotes reflected waves by the boundary, one can obtain the exact dispersion relation $E= J \sum_{j=1}^M\left[\cos \left(k_j\right)-\cosh \phi\right]$ and BAEs \eqref{BAEs_OBC}. Remarkably, both the dispersion relation and BAEs in the OBC are the same as that of the Hermitian counterpart as shown in \eqref{L_OBC_her}. This feature, in fact, can be understood through a imaginary gauge transformation $\hat{S}^{+}_{j}\rightarrow e^{-j\phi}\hat{S}^{+}_{j}$, $\hat{S}^{-}_{j}\rightarrow e^{j\phi}\hat{S}^{-}_{j}$,  $\hat{S}^{z}_{j}\rightarrow \hat{S}^{z}_{j}$ which transforms the non-Hermitian XXZ model \eqref{L_OBC} to its Hermitian counterpart \eqref{L_OBC_her} and meanwhile the wavefunction get an exponential factor $e^{\sum_{j=1}^{M} \phi x_j}$.  Remarkably, in this case the excitation spectrum is gapped, distinct from the PBC case.

Owing to this exponential weight, the spin-up (boson) tends to be more concentrated than Hermitian case resulting in the appearance of the LSE. Fig. \ref{Fig2}(a) shows the density distribution of the steady state for different $\phi$ for the half filling case, from which we can see the skin effect appears. To quantitatively describe the LSE, we introduce the spin imbalance $\mathcal{I}=\frac{N_{\uparrow,r}-N_{\uparrow,l}}{N_{\uparrow,r}+N_{\uparrow,l}}$ where $N_{\uparrow,l}, N_{\uparrow,r}$ ($N_{\uparrow,r}, N_{\downarrow,r}$) is number of the up-spin in the left (right) half of the lattice. In the limit $\phi\rightarrow0$, spins are uniformly distributed  on the lattice. As the growth of  $\phi$,  up-spin particles tend toward occupying part of the lattice near right (left) boundary, and the spin imbalance $\mathcal{I}$ rose from $0$ to $1$ (see Fig. \ref{Fig2}(b)). This means that, in the OBC, the system ultimately relaxes to the steady state with LSE regardless of the choice of initial states. 
Physically, the asymmetric hopping causes a spin current leading to the up-spin congregating near the right boundary.

\section{Solution in generalized boundary conditions}
\label{Sec. GBC}

\begin{figure}[!h]
	\includegraphics[width=8.5cm]{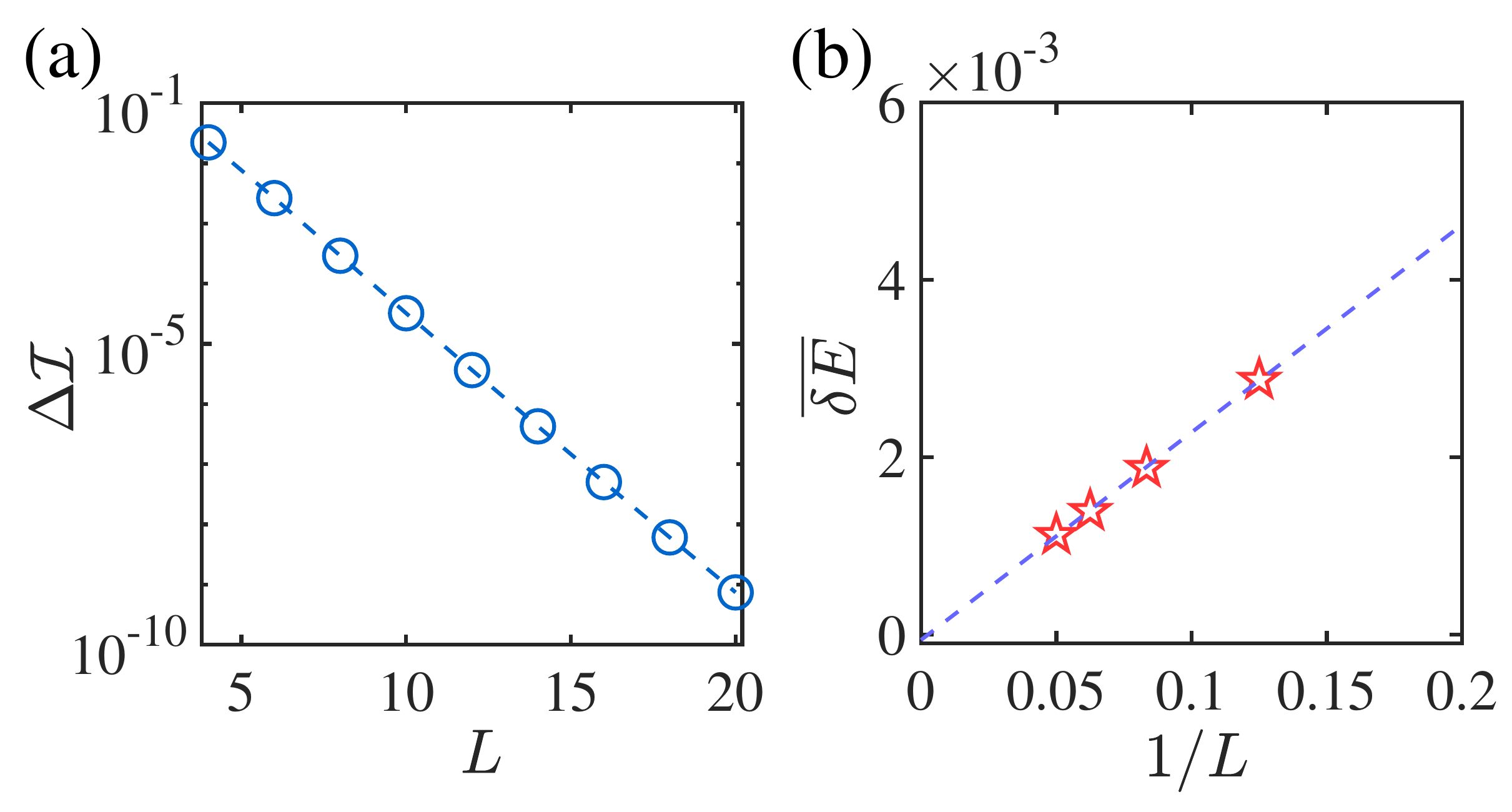}
	\caption{Comparison of the GBC and OBC for imbalance and spectrum. (a) The imbalance deviation $\Delta \mathcal{I}$ between the GBC ($\delta_L\neq0$, $\delta_R=0$) and the OBC case as function of system size $L$. Here we choose $M=L/2$ and $\phi=0.5$, $\delta_L=0.5J_L$. (b) Finite size scaling of the mean energy level difference $\overline{\delta E}$ between the GBC ($\delta_L\neq0$, $\delta_R=0$) and OBC for $M=L/4$.}
	\label{Fig3}
\end{figure}

In this section, we focus on the GBC and investigate the fate of LSE as additional boundary terms are imported. For the sake of convenience, we first study the situation that a count-flow boundary hopping ($\delta_L\neq0$, $\delta_R=0$) is added to the Liouvillian 

\begin{align}
	\hat{\mathscr{L}}_{\text {left}}&=J \sum_{j=1}^{L-1}\Bigg[e^{\phi} \hat{S}_{j+1}^{+} \hat{S}_{j}^{-}+e^{-\phi} \hat{S}_{j}^{+} \hat{S}_{j+1}^{-}\nonumber \\
	&+2 \cosh \phi\left(\hat{S}_{j}^{z} \hat{S}_{j+1}^{z}-\frac{1}{4}\right)\Bigg]+J\sinh \phi\Big(\hat{S}_{L}^{z}- \hat{S}_{1}^{z}\Big)\nonumber\\
	&+\delta_L\left[\hat{S}_{L}^{+} \hat{S}_{1}^{-}
	+\left(\hat{S}_{L}^{z}-\frac{1}{2}\right) \left(\hat{S}_{1}^{z}+\frac{1}{2}\right)\right].
	\label{L_Single_1}
\end{align}
We will see that results in this situation are same as the OBC case in the thermodynamic limit. In order to verify this, we calculate the imbalance difference $\Delta \mathcal{I}=\mathcal{I}_{\text {left}}-\mathcal{I}_{\text {OBC}}$ for the steady state between the Liouvillian \eqref{L_Single_1} and \eqref{L_OBC}. As illustrated in the Fig. \ref{Fig3}(a), the imbalance difference exponentially vanish as increasing of the system size $L$. This means the LSE survives in the presence of count-flow boundary hopping. The reason why the boundary term $\delta_L$ doesn't spoil LSE can be understood through a simple physical picture. 
In large $\phi$ limit, the steady approach to the state $\ket{\downarrow}_1\ket{\downarrow}_2 \cdots . . . \ket{\downarrow}_{L-M}\ket{\uparrow}_{L-M+1}\ket{\uparrow}_{L-M+2} \cdots \ket{\uparrow}_{L}$ in which scenario the boundary hopping is forbidden and meanwhile the right hopping in bulk has no effect on the domain wall. 

For finite $\phi$, we apply the gauge transformation $\hat{S}^{+}_{j}\rightarrow e^{-j\phi}\hat{S}^{+}_{j}$, $\hat{S}^{-}_{j}\rightarrow e^{j\phi}\hat{S}^{-}_{j}$,  $\hat{S}^{z}_{j}\rightarrow \hat{S}^{z}_{j}$ as done in the OBC case.  Then the Liouvillian \eqref{L_Single_1} turns into the Hermitian counterpart of the OBC \eqref{L_OBC_her} plus a boundary term $\delta_L e^{-\phi L}\hat{S}_{L}^{+} \hat{S}_{1}^{-}+\delta_L\left(\hat{S}_{L}^{z}-\frac{1}{2}\right) \left(\hat{S}_{1}^{z}+\frac{1}{2}\right)$ where the effect caused by boundary hopping term $\hat{S}_{L}^{+} \hat{S}_{1}^{-}$ is exponentially small. This conclusion is supported by the result in Fig. \ref{Fig3}(a). Another term 
$\delta_L\left(\hat{S}_{L}^{z}-\frac{1}{2}\right) \left(\hat{S}_{1}^{z}+\frac{1}{2}\right)$ is Hermitian and just serves as boundary potential which doesn't flip boundary spins. This would bring out a $1/L$ correction on the spectrum. To discern this, we calculate the mean energy level difference $\overline{\delta E}=\overline{E}_{\text{left}}-\overline{E}_{\text{OBC}}$ between the Liouvillian \eqref{L_Single_1} and \eqref{L_OBC}. Here the mean energy is defined by 
$\overline{E}=\sum_{j}\overline{E}_j$ with $\overline{E}_j=\frac{E_j-E_{min}}{E_{max}-E_{min}}$ in which $E_{max}$ ($E_{min}$) denotes the maximum (minimum) eigenvalue. The finite size scaling of $\overline{\delta E}$ as plotted in Fig. \ref{Fig3}(b) displays the $1/L$ correction as expected.

Moving on now to consider the more general situation, i.e., neither $\delta_L$ nor $\delta_R$ is zero. First, we find that co-flow boundary hopping strength $\delta_R$ is exponentially amplified ($\delta_Re^{\phi L}\hat{S}_{1}^{+} \hat{S}_{L}^{-}$) if one employs the gauge transformation done above. This implies that we could not extract correct consequences from the gauge transformation, indicating that the wavefunction ansatz in the OBC has failed because it no longer matches boundary conditions induced by $\delta_R$.

%
\begin{figure}[!h]
	\includegraphics[width=8.5cm]{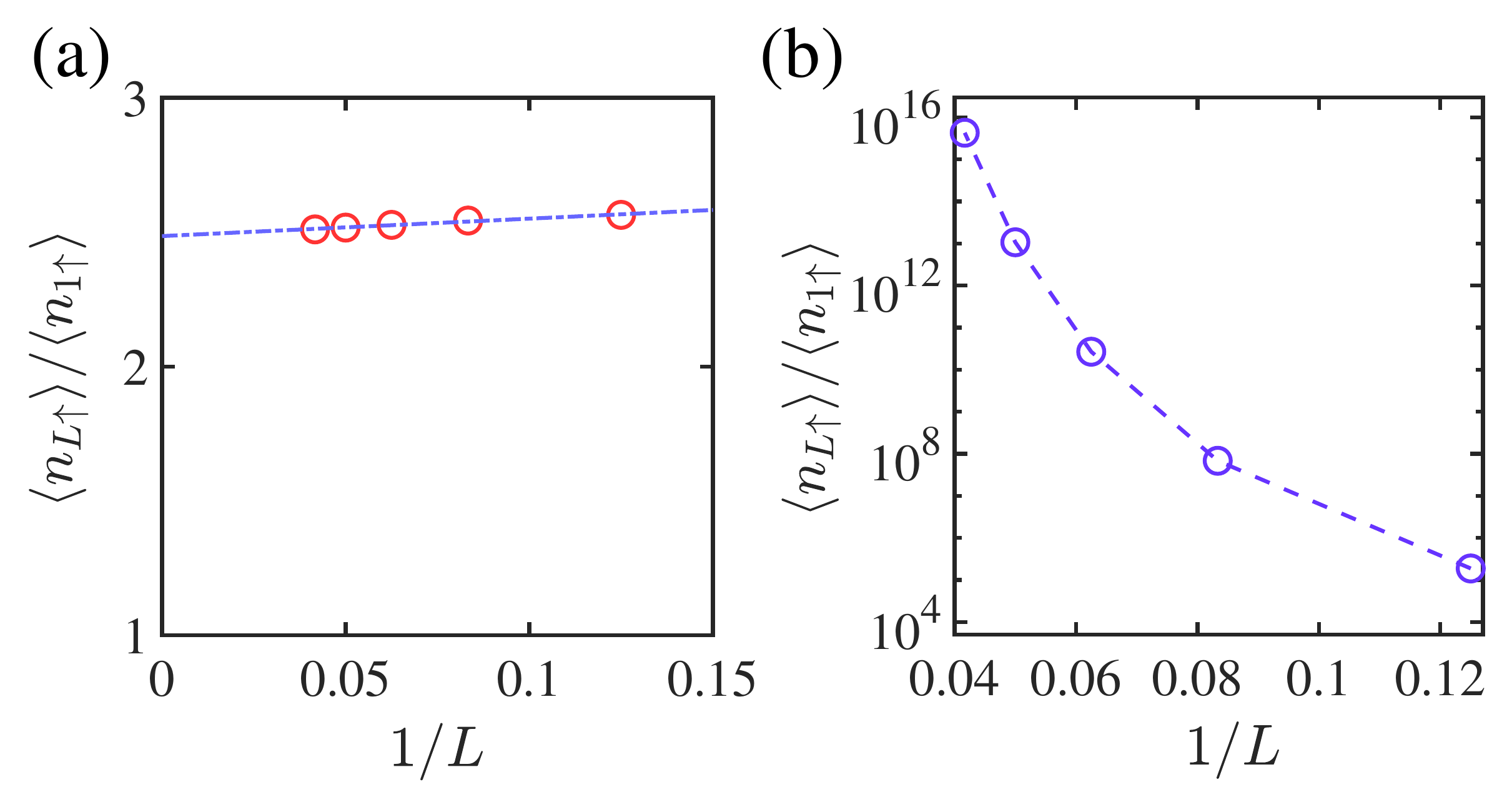}
	\includegraphics[width=8.5cm]{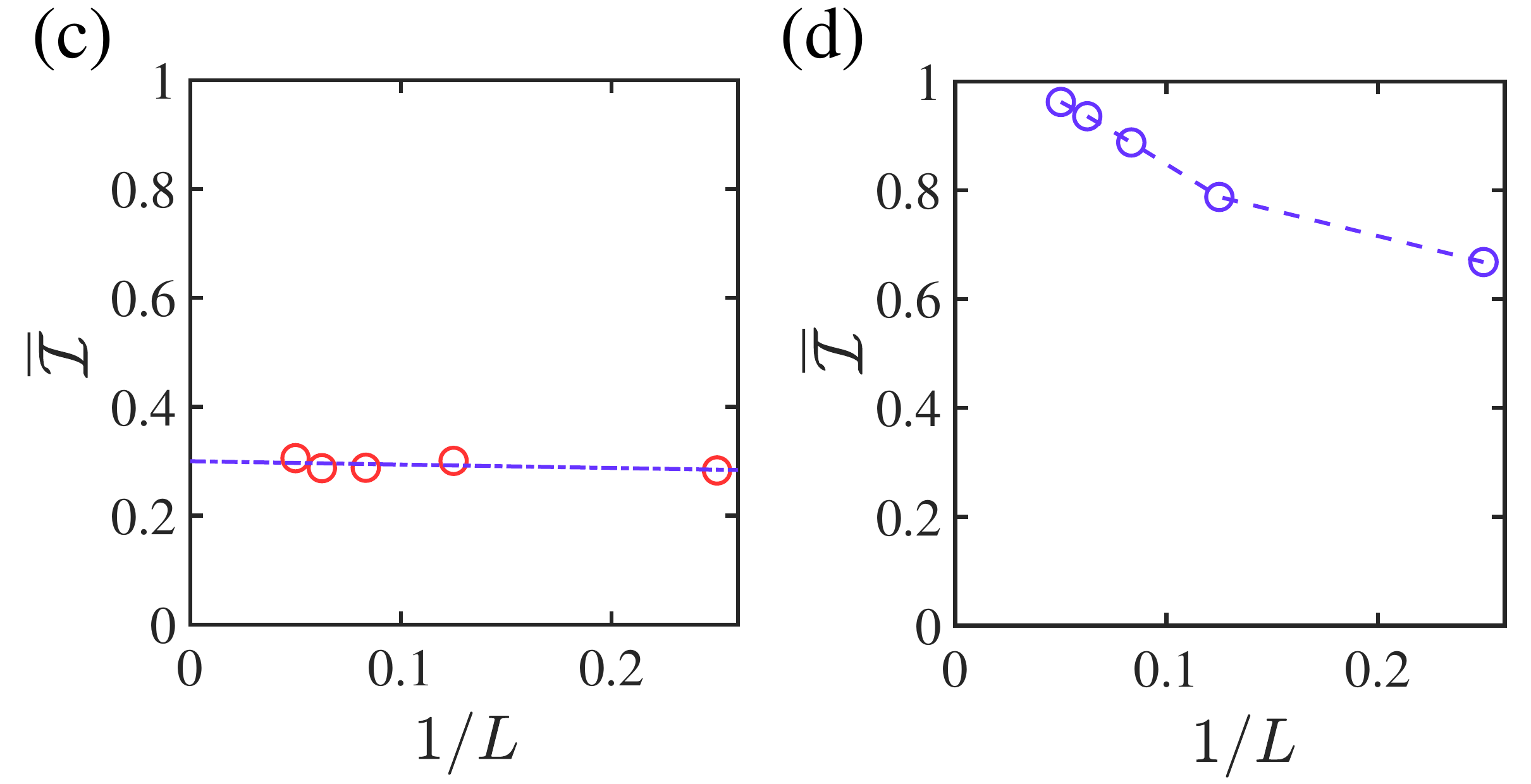}
	\caption{Comparison of occupation ration $\langle n_{L\uparrow}\rangle/\langle n_{1\uparrow}\rangle$ ((a) and (b)) in the steady state and mean imbalance $\overline{\mathcal{I}}$ ((c) and (d))  as function of $1/L$ for two different GBCs. The sub-figures (a), (c)  plot results of the GBC ($\delta_L\neq0$, $\delta_R\neq0$) and the OBC case ($\delta_L\neq0$, $\delta_R=0$) is plotted in (b) and (d). Here we choose $M=L/2$, $\phi=0.5$, $\delta_L=0.5J_L, \delta_R=0.5J_R$ for (a), (c) and  $M=L/2$, $\phi=0.5$, $\delta_L=0.5J_L, \delta_R=0$ for (b), (d).}
	\label{Fig4}
\end{figure}
In fact, as shown in the appendix \ref{App3}, in the thermodynamic limit, the eigenfunction necessarily takes the form of  a PBC-like wavefuction instead of the OBC-like one. Concretely, the PBC-like eigenfunction dismisses the LSE and  consists of modified plane waves $\lambda^{\frac{x_j}{L}}e^{ik_{P_j}x_j}$ instead of original plane waves. The amplitudes of modified plane waves depend on the ratio of $\lambda=Je^{\phi}/\delta_R$, leading to the density distribution at the right boundary ($x_j=L$) only enlarge $\lambda$ times greater than the left boundary ($x_j=1$) which fundamentally differs from the OBC case where the LSE manifests itself in exponential amplifier $e^{\phi L}$ of the density distribution. In order to characterize the  distinction between them, we compute the ratio of up-spin particle number between the right and left boundary $\langle n_{L\uparrow}\rangle/\langle n_{1\uparrow}\rangle$ as illustrated in Fig. \ref{Fig4}(a) and Fig. \ref{Fig4}(b). For the case of GBC $\delta_R=0$, the ratio	 is exponentially divergent due to the LSE while it remains a finite value for the GBC $\delta_R\neq0$ showing the absence of the LSE. Moreover, the distinction is also reflected in the imbalance of eigenstates. Introducing mean imbalance $\overline{\mathcal{I}}=\sum_n \mathcal{I}_{n}/\mathcal{D}$ where $\mathcal{I}_{n}$ denotes the imbalance of $n$-th eigenstate, the significant difference can be visualized from Fig. \ref{Fig4}(c) and Fig. \ref{Fig4}(d). 

\begin{figure}[!h]
	\includegraphics[width=8.5cm]{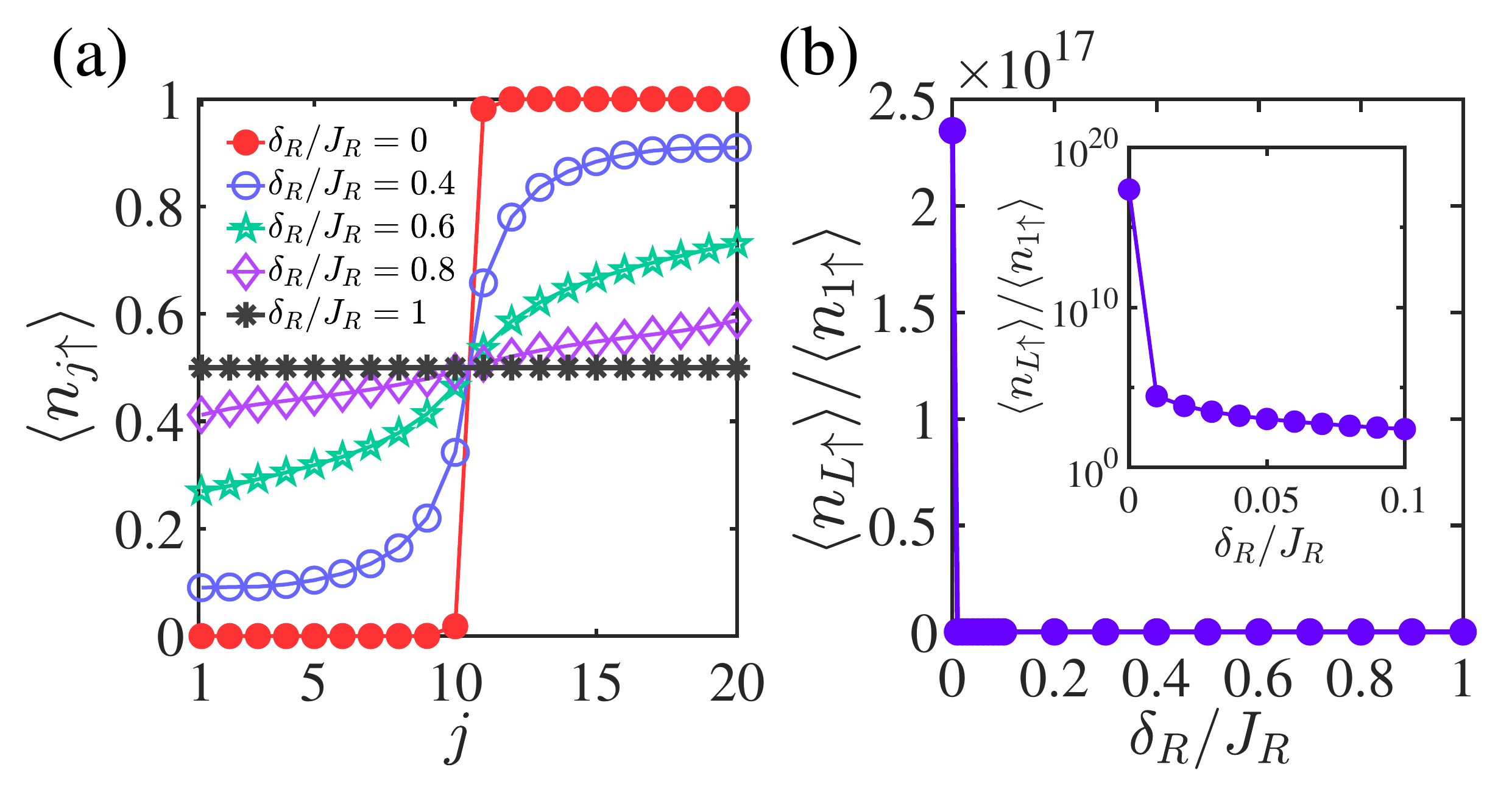}
	\caption{LSE in boundary conditions transitioning from the GBC to PBC at the half filling. (a) The occupation number of  up-spin particles at each site for different $\delta_R$. (b)  The occupation ration $\langle n_{L\uparrow}\rangle/\langle n_{1\uparrow}\rangle$ for different $\delta_R$. The inset shows single logarithm plot where the ration for $\delta_R=0$ is exponential magnitude but follows a power law  for finite $\delta_R\neq=0$.  Here we choose the system length $L=20$,  $M=L/2$, and $\delta_L/J_{L}=1$. The parameter choice $\delta_L/J_{L}=1$ means that the system is in the PBC when $\delta_R/J_{R}=1$.}
	\label{Fig5}
\end{figure}

For the GBC ($\delta_R=0$), $\overline{\mathcal{I}}$ gradually approaches one as the system size grows since eigenstates exhibit the LSE, but it has a nearly constant once a finite $\delta_R\neq0$ involved. That is to say, the LSE is fragile to under the perturbation of co-flow boundary hopping $\delta_R$ but can survive for the counter-flow term $\delta_L$. To see this more clearly, we investigate the parameter variations from the vanishing co-flow boundary coupling $\delta_R/J_{R}=0$ to the full co-flow boundary coupling $\delta_R/J_{R}=1$ with a fixed counter-flow strength $\delta_L/J_{L}=1$. This corresponds to the transition from  GBC with $\delta_R=0,\delta_L\neq0$ to the PBC. Fig. \ref{Fig5}(a) plots the occupation number of up-spin particles at half-filling, which shows that almost all up-spins occupy the right half lattice exhibiting the LSE when $\delta_R=0$. And as co-flow boundary coupling $\delta_R$ increases, the occupation number smooths gradually and eventually reach to discrete uniform distribution at the PBC case  $\delta_L/J_{L}=1$. To examine the fate of LSE further, we calculate the occupation ratio between the right and left boundary $\langle n_{L\uparrow}\rangle/\langle n_{1\uparrow}\rangle$ as shown in Fig.\ref{Fig5}(b). One can see clearly that exponentially divergent ratio $e^{\phi L}$ appear in $\delta_R=0$ which means the LES exists, while the LSE vanishes for any finite co-flow boundary term $\delta_R\neq0$ since the $\langle n_{L\uparrow}\rangle/\langle n_{1\uparrow}\rangle$ just algebraically relates to $J_R/\delta_R$ and toward finite in the thermodynamic limit (see Fig.\ref{Fig4}(a).)

%

\section{Summary and outlook}
\label{Summary}

In summary, through the non-Hermitian XXZ model as an example of exactly solvable many-body Liouvillian superoperators, we detailedly solved the model in various boundary conditions utilizing Bethe ansatz, which enables us to investigate the existence condition of the LSE in different boundary conditions. The main conclusions are summarized as follows. First, for the PBC in which no LSE occurs, there is no phase transition (gap close) in the system. Second, for the OBC, the wavefunction is intrinsically distinct from the PBC one due to the existence of  LSE. We identified the LSE by investigating the density distribution and spin imbalance. We then explore the fate of the LSE in the presence of boundary hoppings. The LSE survives in a kind of GBC with non-zero counter-flow hopping at the boundary because exact solutions, in this case, are equivalent to ones of the OBC in the thermodynamic limit.
In contrast, the LSE will be destroyed immediately once a co-flow hopping appears at the boundary, with roots going back to the similarity to solutions in the PBC. An obvious fact is that boundary conditions would have no consequences on the system in thermodynamic limit. This is indeed true in Hermitian systems. For open quantum systems being intrinsically non-Hermitian, the properties of the system could strongly depend on the boundary condition, even in the thermodynamic limit. This paper is a beneficial endeavor to understand the LSE in exactly solvable open quantum many-body systems. These results add to the expanding field of open quantum systems and could provide a deeper understanding of the non-Hermitian realm. More further studies will be desired to explore the integrability of many-body Liouvillian superoperators, especially under constraint boundary conditions. {Generally, existing literature mainly deals with those non-Hermitian models whose Hermitian parts are integrable. However, a recent study \cite{NH_BHM} found that a particular non-Hermitian Bose-Hubbard model with unidirectional hopping turned out to be integrable, although the Bose-Hubbard model itself is not integrable and cannot be analytically solved. This discovery opens the avenue for further explorations and would produce exciting findings in open many-body quantum systems. 
\\
 
\section*{Acknowledgements}
The work is supported by National Natural Science Foundation of China (Grant No. 12304290 and  No. 12304388), National Key Research and Development Program of China (Grant No. 2022YFA1405800). LP also acknowledges support from the  Fundamental Research Funds for the Central Universities.  \\

\appendix
\section{Exact solution for the non-Hermitian XXZ model in periodic boundary condition} \label{App1}
In this appendix, we solve the non-Hermitian XXZ model in periodic boundary condition (PBC) by means of Bethe ansatz. The model is written as  
\begin{align} 
\hat{\mathscr{L}}=2J \sum_{j=1}^{L}\left[\frac{e^{-\phi}}{2} \hat{S}_j^{+} \hat{S}_{j+1}^{-}+\frac{e^\phi}{2} \hat{S}_{j+1}^{+} \hat{S}_j^{-}+\cosh \phi\left(\hat{S}_j^z \hat{S}_{j+1}^z-\frac{1}{4}\right)\right],\label{NH_XXZ_PBC}
\end{align}
where $\phi>0$ denoting the nonreciprocal spin hopping and PBC is applied ($\hat{S}_{L+1}^{x,y,z}=\hat{S}_{1}^{x,y,z}$). 
Now we show this non-Hermitian XXZ model (\ref{NH_XXZ_PBC}) can be exactly solved by Bethe ansatz. According to the standard procedure of Bethe ansatz, we start from the reference state
\begin{eqnarray}
|\operatorname{Vac}\rangle=|\downarrow \downarrow \cdots \cdot \downarrow\rangle,
\end{eqnarray} 
which satisfies $\hat{\mathscr{L}}|\operatorname{Vac}\rangle=E_0|\operatorname{vac}\rangle$ with $E_0=0$. This means the reference state is the steady state of the Lindbladian. 
Writing eigenfunction in the $S^Z=M-\frac{L}{2}$ sector with the number of down-spin particle $M$ (without loss of generality, we set $M \leqslant \frac{L}{2}$) 
\begin{align} 
|\psi\rangle=\sum_{j=1}^M \sum_{x_j=1}^L \varphi\left(x_1, x_2, \cdots, x_M\right) S_{x_1}^{+} S_{x_2}^{+} \cdots S_{x_M}^{+}|\mathrm{vac}\rangle. \label{BA_wavefun_PBC}
\end{align} 
We restrict the state in the region $1<x_1<x_2<\cdots<x_M<L$ since states in other regions can be obtained by permutation symmetry.

According to the eigen-equation $\hat{\mathscr{L}}|\psi\rangle=E|\psi\rangle$, one can derive the following relation
\begin{widetext}
	\begin{eqnarray}
		J \sum_j\left(1-\delta_{x_j+1, x_{j+1}}\right)\left[\psi\left(x_1, \cdots, x_j+1, x_{j+1}, \cdots, x_M\right)e^{-\phi}+\psi\left(x_1, \cdots, x_j, x_{j+1}-1, \cdots, x_M\right) e^{\phi}\right]\nonumber \\
		+\left[E_0-E- 2J \cosh \phi\left(M-\sum_j \delta_{x_j+1, x_{j+1}}\right) \right]\varphi\left(x_1, x_2, \cdots, x_M\right)=0.
	\end{eqnarray}
\end{widetext}
Constructing the many-body wavefunction by means of the following Bethe ansatz form
\begin{align} 
\varphi\left(x_1, x_2, \cdots, x_M\right)=\sum_{\bf P} A_{\bf P} \exp \left(i \sum_{j=1}^M k_{p_j} x_j\right),
\end{align}
where ${\bf P}=\left(p_1, p_2, \cdots, p_M\right)$ is a permutation of $1,2, \cdots, M$, then yields the eigenvalue
\begin{align} 
E=E_0+2J \sum_{j=1}^M\left[\cos \left(k_j+i\phi\right)-\cosh \phi\right],
\end{align} 

and the Bethe ansatz equations 
\begin{widetext}
\begin{align} 
	\exp\left(i k_j L\right)=(-1)^{M-1} \prod_{l \neq j}^{M} \frac{\exp \left[i\left(k_j+k_l\right)-2\phi\right]+1-2 \cosh \phi \cdot \exp \left(i k_j-\phi\right)}{\exp \left[i\left(k_j+k_l\right)-2\phi\right]+1-2 \cosh \phi \cdot \exp \left(i k_l-\phi\right)},\label{App_BAEs1}
\end{align} 
\end{widetext}
via the PBC $\varphi\left(x_1, x_2, \cdots, x_M\right)=\varphi\left( x_2, \cdots, x_M,x_1+L\right)$. For the sake of convenience, introducing rapidity parameters $\left\{\lambda_j\right\}$
\begin{align} 
\exp \left(i k_j-\phi\right)=-\frac{\sin \left[\phi\left(\lambda_j+i\right) / 2\right]}{\sin \left[\phi\left(\lambda_j-i\right) / 2\right]},
\end{align} 
then \eqref{App_BAEs1} becomes
\begin{align} 
\left[\frac{\sin \frac{\phi}{2}\left(\lambda_j+i\right)}{\sin \frac{\phi}{2}\left(\lambda_j-i\right)}\right]^L e^{\phi L}=\prod_{l \neq j}^{M} \frac{\sin \frac{\phi}{2}\left(\lambda_j-\lambda_l+2 i\right)}{\sin \frac{\phi}{2}\left(\lambda_j-\lambda_l-2 i\right)}.
\end{align} 
Taking the logarithm of above equations, we can obtain
\begin{align} 
L \theta_1\left(\lambda_j\right)=2 \pi I_j+i \phi L+\sum_{l \neq j}^M \theta_2\left(\lambda_j-\lambda_l\right),\label{logBAEs}
\end{align}
where $\theta_n(\lambda)=2 \arctan \left[\tan \left(\frac{\phi \lambda}{2}\right)\operatorname{coth}\left(\frac{n \phi}{2}\right) \right]$.
For the ground state $\left(M=\frac{L}{2}\right)$, we have $I_j=-(\frac{L}{2}-1)/2,-(\frac{L}{2}-1)/2+1, \cdots,(\frac{L}{2}-1)/2-1,  (\frac{L}{2}-1)/2$.
In the thermodynamic limit $L \rightarrow \infty$,  Eq.\eqref{logBAEs} becomes

\begin{align} 
2 \theta_1(\lambda)=2 \pi \int^\lambda \sigma\left(\lambda^{\prime}\right) d \lambda^{\prime}+i \phi+2 \int_\mathcal{C} \theta_2(\lambda-\lambda^{\prime}) \sigma(\lambda^{\prime}) d \lambda^{\prime} \label{theta1}
\end{align}
where $\mathcal{C}$ is the rapidity path in complex plain. Since $\theta_1$ is a function functions with a periods of $\frac{2\pi}{\phi}$, one can restrict the domain of definition to
$ \frac{\pi}{\phi} \leqslant \operatorname{Re} \lambda \leqslant \frac{\pi}{\phi}$.
Differentiating  \eqref{theta1}, the distribution $\sigma(\lambda)$ satisfies the equation
\begin{align} 
\frac{\phi \sinh \phi}{\cosh \phi-\cos (\phi \lambda)}&=2 \pi \sigma(\lambda)\nonumber\\
&+\int_{\mathcal{C}} \frac{\phi \sinh (2 \phi)}{\cosh (2 \phi)-\cos [\phi(\lambda-\lambda^{\prime})]} \sigma(\lambda^{\prime}) d \lambda^{\prime},
\end{align} 
from which one can find two poles $\Rightarrow \lambda-\lambda'= \pm 2 i$ in the integrand. 
If $-1<\operatorname{Im} \lambda<1$ and $-1<\operatorname{Im} \lambda'<1$, then the integral path dose not enclose poles which means the path can be continuously deformed to real axis. 
As $\phi$ grows, the path $\mathcal{C}$ gradually extends to the complex plane and then touch poles when $\phi=\phi_c$ in which case Eq. \eqref{theta1} diverges and the mass gap is closed \cite{NH_XXZ}.

Next we calculate the critical value $\phi_c$. When $\phi<\phi_c$, the poles are not enclosed by curve $\mathcal{C}$ and the integral path can be continuously deformed to the real axis in which case 
\begin{align}
\frac{\phi \sinh \phi}{\cosh \phi-\cos (\phi \lambda)}=2 \pi \sigma(\lambda)+\int_{-\frac{\pi}{\phi}}^{\frac{\pi}{\phi}} \frac{\phi \sinh (2 \phi)}{\cosh (2 \phi)-\cos [\phi(\lambda-1)]} \sigma(\Lambda) d \lambda . \label{BAEs_sigma}
\end{align}
We can solve $\sigma(\lambda)$ by performing Fourier transformation of Eq.\eqref{BAEs_sigma}
 
\begin{align}
\sigma(\lambda)=\sum_{m=-\infty}^{\infty} \frac{e^{-i m \phi \lambda}}{2 \cosh (m \phi)} .\label{sigma_Fourier}
\end{align}
After substitute Eq. \eqref{sigma_Fourier}  to Eq. \eqref{theta1} and consider the following expansion

\begin{align}
\theta_1(\lambda)&=\phi \lambda-\sum_{m \neq 0} \frac{\exp (-i m \phi \lambda-\phi|n|)}{i m \phi} \nonumber \\ \theta_2(\lambda)&=\phi \lambda-\sum_{m \neq 0} \frac{\exp (-i m \phi \lambda-2 \phi|n|)}{i m \phi},
\end{align}
 
we can obtain the equation
 
\begin{align}
& \phi=2 \pi i Z(\lambda)+\frac{\phi}{2} b-\frac{i}{2} \phi \lambda+\sum_{m \neq 0} \frac{e^{-i m \phi}}{2 m \cosh (m \phi)}\nonumber \\
& +\sum_{m \neq 0}(-1)^m \frac{e^{m \phi b}}{2 m \cosh (m \phi)}.  
\end{align}
 
The critical value $\phi_c$ is determined by $\lambda= \pm \frac{\pi}{\phi}+i$

\begin{align}
	\phi=\phi+\sum_{m=1}^{\infty} \frac{(-1)^m \tanh (m \phi)}{m}, \label{phi_c}
\end{align}
from which we find there is only a trivial solution
$\phi_c=0$. That means no gap close happens.


\section{Exact solution for non-Hermitian XXZ model in open boundary conditions} \label{App2} 
In this appendix, we discuss the non-Hermitian XXZ model in OBC, which exhibits the skin effect. In the OBC, the Liouvillian is expressed as
\begin{align}
\hat{\mathscr{L}}_{\text {OBC}}&=J \sum_{j=1}^{L-1}\Bigg[e^{\phi} \hat{S}_{j+1}^{+} \hat{S}_{j}^{-}+e^{-\phi} \hat{S}_{j}^{+} \hat{S}_{j+1}^{-}\nonumber \\
&+2 \cosh \phi\left(\hat{S}_{j}^{z} \hat{S}_{j+1}^{z}-\frac{1}{4}\right)\Bigg]+J\sinh \phi\Big(\hat{S}_{L}^{z}- \hat{S}_{1}^{z}\Big).\label{L_OBC_1}
\end{align}
For simplicity,  we first discuss the limit $\phi \rightarrow \infty$, in which the Liouvillian can be reduced to the following form (in unit of $e^{\phi}$)

\begin{align}
\hat{\mathscr{L}}_{\text {OBC}}=J\sum_{j=1}^{L-1}\left[S_{j+1}^{+} \hat{S}_{j}^{-}+\left(\hat{S}_{j}^{z} \hat{S}_{j+1}^{z}-\frac{1}{4}\right)\right] +\frac{J}{2}\Big(\hat{S}_{L}^{z}- \hat{S}_{1}^{z}\Big).\label{L_OBC_2}
\end{align}

Notably, the matrix representation of Liouvillian \eqref{L_OBC_2} is a triangular matrix in the basis
\begin{align}
	\Bigg\{\ket{\downarrow \cdots\downarrow\downarrow\uparrow \uparrow\cdots  \uparrow},\ket{\downarrow \cdots\downarrow\uparrow\downarrow \uparrow\cdots  \uparrow},\ket{\uparrow\uparrow  \cdots\uparrow\downarrow\uparrow\downarrow \downarrow\cdots  \downarrow},\nonumber \\ \cdots,\ket{\uparrow\uparrow  \cdots\uparrow\downarrow\uparrow\downarrow \downarrow\cdots  \downarrow},\ket{\uparrow\uparrow \cdots \uparrow\downarrow\downarrow \cdots  \downarrow}\Bigg\}, 
\end{align} in which case it can easily be diagonalized.
One can find that the state $\ket{\downarrow}_1\ket{\downarrow}_2 \cdots . . . \ket{\downarrow}_{M}\ket{\uparrow}_{M+1}\ket{\uparrow}_{M+2} \cdots . . . \ket{\uparrow}_{L}$ where a spin domain-wal plateau emerges is exact eigenstate. This eigenstate is also a steady state because it has zero eigenvalue. This means that the system relaxes eventually to the skin mode where all up spins (bosons) arrange right sites.
For the finite $\phi$,  the Liouvillian \eqref{L_OBC_1} is solvable with Bethe ansatz. Here it should be emphasized that the Liouvillian \eqref{L_OBC_1} can not be solved simply in terms of  traditional Bethe ansatz for Hermitian case in OBC, i.e., $\varphi\left(x_1, x_2, \cdots, x_M\right)=\sum_{\bf P, r_1,\cdots,r_M} A_{\bf P}(r_1,r_2,\cdots,r_M)  \exp \left(\sum_{j=1}^M  
i k_{p_j} x_j\right)$. One should adopt to the following ansatz 
\begin{align} 
\varphi\left(x_1, x_2, \cdots, x_M\right)&=\sum_{\bf P, r_1,\cdots,r_M} A_{\bf P}(r_1,r_2,\cdots,r_M)\nonumber \\
&\times\exp \left[\sum_{j=1}^M\left(i r_ j k_{p_j} x_j+\phi x_j\right)\right], \label{OBC_wavefun}
\end{align}
where $r_ j =\pm 1$ denotes the plane wave of the $j$-th
particle traveling wave toward left ($r_ j =-1$) and right ($r_ j =1$).  In this way, after some calculations, one can obtain the following Bethe ansatz equations
\begin{align} 
	\mathrm{e}^{i 2(L-1) k_j} &\frac{\left(\mathrm{e}^{i k_j}-\sinh\phi-\cosh\phi\right)\left(\mathrm{e}^{i k_j}+\sinh\phi-\cosh\phi\right)}{\left(\mathrm{e}^{-ik_j}-\sinh\phi-\cosh\phi\right)\left(\mathrm{e}^{-\mathrm{i} k_j}+\sinh\phi-\cosh\phi\right)}\nonumber\\
	&=\prod_{l \neq j}^{M} \frac{S\left(-k_j, k_l\right)S\left(k_l, k_j\right)}{S\left(k_j, k_l\right)S\left(k_l, -k_j\right)}.\label{BAEs_OBC}
\end{align} 
with $S\left(k_j, k_l\right)=1-2 \cosh\phi\mathrm{e}^{i k_l}+\mathrm{e}^{\mathrm{i}\left(k_j+k_l\right)}$ and the associated eigenvalues 
\begin{align} 
	E= 2J \sum_{j=1}^M\left[\cos \left(k_j\right)-\cosh \phi\right].\label{Ener_OBC}
\end{align} 

One can find that the BAEs and eigenvalues are identical to the Hermitian XXZ model in OBC, and the only difference between them exists in the wavefunction where $\phi$ will appear in the non-Hermitian system as shown in Eq.\eqref{OBC_wavefun}.
In order to understand this, we employ a gauge transformation $\hat{S}^{+}_{j}\rightarrow e^{-j\phi}\hat{S}^{+}_{j}$, $\hat{S}^{-}_{j}\rightarrow e^{j\phi}\hat{S}^{-}_{j}$,  $\hat{S}^{z}_{j}\rightarrow \hat{S}^{z}_{j}$ to eliminate the unequal hopping which reproduces the Hermitian counterpart
\begin{align}
	\hat{\mathscr{L}}_{\text {Hermitian}}&=J \sum_{j=1}^{L-1}\Bigg[\hat{S}_{j+1}^{+} \hat{S}_{j}^{-}+\hat{S}_{j}^{+} \hat{S}_{j+1}^{-}\nonumber \\
	&+2 \cosh \phi\left(\hat{S}_{j}^{z} \hat{S}_{j+1}^{z}-\frac{1}{4}\right)\Bigg]+J\sinh \phi\Big(\hat{S}_{L}^{z}- \hat{S}_{1}^{z}\Big).\label{L_OBC_her}
\end{align}
Thus, it is straightforward that for the Hermitian XXZ model \eqref{L_OBC_her}, BAEs and eigenvalues are given by the 
Eq.\eqref{BAEs_OBC} and Eq.\eqref{Ener_OBC}. 
But the wavefunction to be changed correspondingly, $\varphi\left(x_1, x_2, \cdots, x_M\right)\rightarrow\varphi\left(x_1, x_2, \cdots, x_M\right)e^{\sum_{j=1}^{M}\phi x_j}$ which is nothing but the ansatz \eqref{OBC_wavefun}. 
This implies that there exists NHSE in OBC, because tthe solution of ${k_j}$ is identical to Hermitian case which is determined by BAEs \eqref{L_OBC_her} and meanwhile the wavefunction capture exponential factor.\\

\section{Exact solution for non-Hermitian XXZ model in generalized boundary conditions} \label{App3} 

After solving the non-Hermitian XXZ model in PBC and OBC,  we discuss other situations that one unidirectional hopping term appears at the boundary. We first investigate the following situation where the Liouvillian is expressed as  

\begin{align}
	\hat{\mathscr{L}}_{\text {left}}&=J \sum_{j=1}^{L-1}\Bigg[e^{\phi} \hat{S}_{j+1}^{+} \hat{S}_{j}^{-}+e^{-\phi} \hat{S}_{j}^{+} \hat{S}_{j+1}^{-}\nonumber \\
	&+2 \cosh \phi\left(\hat{S}_{j}^{z} \hat{S}_{j+1}^{z}-\frac{1}{4}\right)\Bigg]+J\sinh \phi\Big(\hat{S}_{L}^{z}- \hat{S}_{1}^{z}\Big)\nonumber\\
	&+\delta_L\left[\hat{S}_{L}^{+} \hat{S}_{1}^{-}
	+\left(\hat{S}_{L}^{z}-\frac{1}{2}\right) \left(\hat{S}_{1}^{z}+\frac{1}{2}\right)\right].
	\label{H_Single_1}
\end{align}
Here $\delta_L\hat{S}_{L}^{+} \hat{S}_{1}^{-}$ denotes the unidirectional hopping from the left to the right boundary. 

Starting from the OBC ansatz \eqref{OBC_wavefun}, Since the boundary hopping term has no impact on the bulk spins, the eigenenergy is same as the OBC case  \eqref{Ener_OBC}. However,  it is expected that the boundary equations will be modified and a straightforward calculation derives 
\begin{widetext}
\begin{subequations}	
\begin{align}
	& J \sum_{j=2}^{M}\sum_{\delta= \pm 1} \left(1-\delta_{x_j+\delta, x_{j+\delta}}\right)\phi\left(1, \cdots, x_j+\delta, \cdots, x_M\right)e^{-\delta \phi}-
	\left[E+ 2J \cosh \phi\left(M-1/2-\sum_{j=2}^{M}\delta_{x_j+1, x_{j+1}}\right)\right] \varphi\left(1, x_2, \cdots, x_M\right)\nonumber \\
	&+Je^{-\phi}\varphi\left(2, x_2\cdots, x_j, \cdots, x_M\right)-(J\sinh \phi+\delta_L)~\varphi\left(1,x_2,\cdots, x_j, \cdots, x_{M-1},x_M\right)=0, 
	\label{left_boundary}\\
	& J \sum_{j=1}^{M-1}\sum_{\delta= \pm 1} \left(1-\delta_{x_j+\delta, x_{j+\delta}}\right)\phi\left(x_1, \cdots, x_j+\delta, \cdots, L\right)e^{-\delta \phi}-\left[E+ 2J \cosh \phi\left(M-1/2-\sum_{j=1}^{M-1}\delta_{x_j+1, x_{j+1}}\right)\right] \varphi\left(x_1, x_2, \cdots, L\right)\nonumber \\
	&+Je^{\phi}\sum_{j=1}^{M-1}\varphi\left(x_1, \cdots, x_j, \cdots, L-1\right)+\uwave{\delta_L\varphi\left(1,x_1,\cdots, x_j, \cdots, x_{M-1}\right)}+J\sinh \phi~\varphi\left(x_1,\cdots, x_j, \cdots, x_{M-1},L\right)=0, 
	\label{right_boundary}
\end{align}
\end{subequations}
\end{widetext}
where the term marked with a wavy line originates from the boundary hopping term. From the OBC ansatz \eqref{OBC_wavefun}, one can find the boundary hopping term is exponentially smaller than other terms since they are 
amplified with a factor $e^{L\phi}$. This means that the boundary term vanishes in thermodynamic limit which indicates solutions are identical to the OBC case. Hence, it is conceivably concluded that the system exhibits NHSE where up-spin particles accumulate toward the right boundary. An alternative explanation for NHSE exists in this system is that in large $\phi$ limit,  similar to the OBC case,  the state $\ket{\downarrow}_1\ket{\downarrow}_2 \cdots . . . \ket{\downarrow}_{M}\ket{\uparrow}_{M+1}\ket{\uparrow}_{M+2} \cdots . . . \ket{\uparrow}_{L}$ is zero-energy eigenstate because the boundary hopping term has no effect on it. 

We now turn to discuss the generalized boundary condition with $\delta_L\neq0$ and $\delta_R\neq0$ as shown in Eq.\eqref{L_GBZ1}.
In this case, the boundary equations are given by

\begin{widetext}
	\begin{subequations}	
		\begin{align}
& J \sum_{j=2}^{M}\sum_{\delta= \pm 1} \left(1-\delta_{x_j+\delta, x_{j+\delta}}\right)\varphi\left(1, \cdots, x_j+\delta, \cdots, x_M\right)e^{-\delta \phi}-
\left[E+ 2J \cosh \phi\left(M-1/2-\sum_{j=2}^{M}\delta_{x_j+1, x_{j+1}}\right)\right] \varphi\left(1, x_2, \cdots, x_M\right)\nonumber \\
&+Je^{-\phi}\varphi\left(2, x_2\cdots, x_j, \cdots, x_M\right)+\uwave{\delta_R\varphi\left(x_2,\cdots, x_j, \cdots, x_{M},L\right)}-(J\sinh \phi+\delta_L) \varphi\left(1,x_2,\cdots, x_j, \cdots, x_{M-1},x_M\right)=0, 
			\label{left_boundary2}\\
			& J \sum_{j=1}^{M-1}\sum_{\delta= \pm 1} \left(1-\delta_{x_j+\delta, x_{j+\delta}}\right)\varphi\left(x_1, \cdots, x_j+\delta, \cdots, L\right)e^{-\delta \phi}-\left[E+ 2J \cosh \phi\left(M-1/2-\sum_{j=1}^{M-1}\delta_{x_j+1, x_{j+1}}\right)\right] \varphi\left(x_1, x_2, \cdots, L\right)\nonumber \\
			&+Je^{\phi}~\varphi\left(x_1, \cdots, x_j, \cdots, L-1\right)+\delta_L\varphi\left(1,x_1,\cdots, x_j, \cdots, x_{M-1}\right)+(J\sinh \phi-\delta_R)~\varphi\left(x_1,\cdots, x_j, \cdots, x_{M-1},L\right)=0,  
			\label{right_boundary2}
		\end{align}
	\end{subequations}
\end{widetext}
where the term marked with a wavy line comes from the boundary hopping.  If we still use the OBC ansatz, the boundary term exponentially grows with system size, which far outweigh the 
other terms. This is unreasonable because the equation is dominated only by a boundary term  no matter how small for $\delta_R$.  Therefore, a alternative ansatz should be employed. Suppose the quasi-momentum takes complex value and we perform a substitution $k_j\rightarrow k_j+i\phi$. It will be found that this substitution will cancel the exponential divergence in the boundary term. The boundary equation \eqref{right_boundary2} gives rise to the relation $\frac{A_{\bf P}(r_1,r_2,\cdots,-)}{A_{\bf P}(r_1,r_2,\cdots,+)}\sim e^{-2\phi L}$
which means $\frac{A_{\bf P}(r_1,r_2,\cdots,-)}{A_{\bf P}(r_1,r_2,\cdots,+)}\rightarrow0$ in the large $L$ limit. Moreover,  when $x_{j+1}=x_j+1$, the contact condition 
gives 
\begin{align}
\frac{A_{ P_1,\cdots,P_j,P_{j+1},\dots,P_M}(r_1,\cdots,r_{j},r_{j+1},\cdots,r_M)}{A_{ P_1,\cdots,P_{j+1},P_j,\dots,P_M}(r_1,\cdots,r_{j+1},r_j,\cdots,r_M)}=-\frac{S(k_{P_{j+1}},k_{P_j})}{S(k_{P_j},k_{P_{j+1}})}.\label{A_A2}
\end{align}
Combining Eq.\eqref{A_A2} with the condition $\frac{A_{\bf P}(r_1,r_2,\cdots,-)}{A_{\bf P}(r_1,r_2,\cdots,+)}\rightarrow0$ produce that all coefficients $A_{\bf P}(r_1,r_2,\cdots,r_M)$ for any $r_j=-1$ are zero. The vanishing left moving wave indicate a no-reflecting boundary condition in the thermodynamic limit.  As a result, we try modified PBC wavefucntion by a substitution into the PBC wavefcuntion  $e^{ik_{P_j}x_j}\rightarrow \lambda^{\frac{x_j}{L}}e^{ik_{P_j}x_j}$ with $\lambda=\frac{Je^{\phi}}{\delta_R}$, and hence $\varphi\left(x_1, x_2, \cdots, x_M\right)=\sum_{\bf P} A_{\bf P}\exp \left(i \sum_{j=1}^M k_{p_j} x_j+x_j\log\lambda \right)$. After this substitution, the coefficient $\delta_L$  of the term with wavy line in \eqref{left_boundary2} transforms to $J_R$ which is same as the boundary hopping in the PBC. In fact, from boundary condition \eqref{left_boundary2}, using the modified PBC wavefunction, one can derive a PBC-like BAEs
\begin{widetext}
	\begin{align} 
		\exp\left(i k_j L\right)= (-1)^{M-1}\kappa\prod_{l \neq j}^{M} \frac{\exp \left[i\left(k_j+k_l\right)-2\phi\right]+1-2 \cosh \phi \cdot \exp \left(i k_j-\phi\right)}{\exp \left[i\left(k_j+k_l\right)-2\phi\right]+1-2 \cosh \phi \cdot \exp \left(i k_l-\phi\right)}.\label{BAEs_deltaR}
	\end{align} 
\end{widetext}
We can see clearly that BAEs \eqref{BAEs_deltaR} are same as the ones in PBC case up to a coefficient $\kappa=1-\frac{(J_L-\delta_L)(\frac{J_R}{\delta_R})^{1/L}~e^{ik_j}}{J_R}$ 
which serves as a boundary term. However, the boundary term just produce a $1/L$ correction which vanishes in the large $L$ limit which means no LSE exists. 

\end{document}